\documentclass[12pt]{article}
\usepackage{graphicx}
\begin{document}

\centerline{\large \bf Scaling effects in the Penna ageing model}

\bigskip
\centerline{\large A. {\L}aszkiewicz, S. Cebrat and D. Stauffer*}

\bigskip

\noindent
Department of Genomics, Institute of Genetics and Microbiology, 
University of Wroc{\l}aw, 

\noindent
ul. Przybyszewskiego 63/77, PL-54148 Wroc{\l}aw, 
Poland, cebrat@microb.uni.wroc.pl \\

\noindent
*Institute for Theoretical Physics, Cologne University, D-50923 K\"oln,
Euroland,  

\noindent
stauffer@thp.uni-koeln.de 

\noindent
* To whom all correspondence should be sent.

\bigskip
Abstract:
We have analysed the possibility of scaling the sexual Penna ageing model. 
Assuming that the number of genes expressed before the reproduction age grows 
linearly with the genome size and that the mutation rate per genome and 
generation is constant, we have found that the fraction of defective genes 
expressed before the minimum reproduction age drops with the genome size, while 
the number of defective genes eliminated by the genetic death grows with 
genome size. Thus, the evolutionary costs decrease with enlarging the genome. 
After rescaling the time scale according to the mutational clock, age 
distributions of populations do not depend on the genome size. Nevertheless, 
enlarging the genome increases the reproduction potential of populations. 


\section{Introduction}

The Penna ageing model \cite{Penna} is the most often used Monte Carlo model for
simulating the dynamics of age-structured populations. Properly rescaled and 
interpreted simulation results mimic very well the age distribution of real 
populations, including the human one \cite{Ewa}. Changing only one parameter 
describing the relations between the genotype (phenotype) and the environment, 
it is possible to simulate the changes of the age distribution of human 
populations during the last centuries and predict the human life expectancy in 
the future \cite{Aga1}, \cite{Aga2}. One of the most important problems 
concerning this model is the question how the genome size in the model 
influences the results of simulations. Thus far, the genome size implemented in 
the sexual model as two bit-strings, with relatively low number of bits (genes) 
(from 8 up to around 1000 bits), does not correspond to the real sizes of
natural genomes, which in the case of higher eukaryotes are of the order of 
dozens of thousands of genes. Our question was if it is possible to generate in 
simulations populations with similar age distributions independently of the 
genome size. This problem was already addressed by \cite{Malarz} without 
conclusive results and recently by \cite{Jorge} whose results suggested roughly the 
scaling properties of the model. In our analyses we have tried to show how the 
parameters describing the simulated populations change depending on the length 
of the genomes.

\section{Model}

In the Penna sexual model each individual is represented by its genome composed 
of two bit-strings --- each $L$ bits long. Bits (genes) are switched on in pairs, 
consecutively, one pair during each Monte Carlo step. If a bit is set for 0 it 
means that it is functional, if it is set for 1, it is defective. Defective 
genes are recessive --- both bits (alleles) in the same locus have to be defective 
to determine the defective phenotype. After a defined number of steps, and 
switching on the corresponding pairs of bits, the individual reaches the 
minimum
reproduction age $R$. An individual of reproductive age produces 
gametes by crossing over its parental bit-strings at a randomly chosen point 
with a declared probability $C$. One mutation is added to the recombined strings
in one randomly chosen locus. If the bit chosen for mutation is already set to 
$1$ it stays $1$, so there are no reversions. A gamete produced by a female is 
joined by another one produced by a male and offspring is born. Each female 
produces an offspring with a declared probability during each MC step. The sex 
of the newborn is randomly chosen with an equal probability for male and female.
If, after switching on bits in consecutive loci, the declared number $T$ of 
defective phenotypes is reached --- the individual dies because of its genetic 
status. To avoid 
unlimited growth of the population, the Verhulst factor  $V$ is introduced: 
${V=1-N_{t}/N_{max}}$ where ${N_{max}}$ --- the maximum 
population size --- is often called the capacity of the environment, and $N_{t}$
is the current population size. For each zygote a random number between $0$ and 
$1$ is generated and if it is greater than $V$, the zygote 
dies. Thus, there could be three different causes of death:
--- genetic death caused by reaching the threshold $T$ of the number of expressed 
deleterious genes;
--- random death because of overcrowded environment; 
--- death at maximum age which corresponds to the total number of bits in the 
bit-string.
In practice, organisms do not reach the maximum allowable age determined by the 
length of the bit-strings. Since random deaths caused by the Verhulst factor could 
happen only at birth, the only deaths observed in the evolving populations are 
genetic deaths caused by surpassing the declared threshold $T$. 
Summing up, there are only a few parameters crucial for the final state of 
the population simulated by the Penna model:
$T$ - the upper limit of expressed phenotypic defects, at which an individual 
dies;
$R$ - minimum reproduction age;
$B$ - birth rate, the number of offspring produced by each female at 
reproduction age at each time step;
$M$ - mutation rate, the number of new mutations introduced into the haploid 
genome during gamete production;
$C$ - the probability of cross-over between parental haplotypes during gamete 
production or the number of cross-overs.

\section{Scaling experiments}

We have performed three different sets of simulations (parameters shown in Table
1). Note that the cross-over frequency per bit was constant for all simulations 
and no dominant loci were declared.

\medskip
a) $M$ per bit and  $R$ constant.

In the first series, the minimum reproduction age ($8$)  and $M=1$ per bit was 
constant. Under such parameters, in equilibrium, in the population with the 
shortest genome ($L=32$), all bits beyond the $15th$ turned out to be set to 
$1$. It means 
that the rest of the genome is dispensable. Since in the simulations with larger
genomes the parameters in the first part (call it the  monomer genome) are the 
same, the results of simulations are also the same. All genes (and further 
monomers) expressed after the first $15$ loci are dispensable and set for $1$. 
In fact only the first "monomer genome" is responsible for the age structure of 
populations. The results are trivial. This is not a scaling.  

\medskip
b) $M$ per bit constant, $R$ proportional to the genome size.

In the second series, where the mutational pressure per bit was constant and the
minimum reproduction age was proportional to the genome size, the mutational 
pressure exerted on the genes expressed before reproduction grew and as a result
the populations with longer genomes died out. The conclusion seems to be 
obvious - species with larger genomes require higher fidelity of replication. 

\medskip
c) $M$ per genome constant, $R$ proportional to genome size.

In the third series of simulations the mutation rate per genome and generation 
was constant. The constant mutational pressure per genome seems to be 
biologically legitimate. Besides the conclusions from the second series of 
simulations, many experiments on living systems \cite{Drake} as well as 
theoretical considerations \cite{Azbel} indicate that the mutational pressure is
of the order of one mutation per genome replication independently of the size of
the genome. Thus, the time of the simulations may be measured in MC steps (each 
step corresponding to a single bit) or by the "mutational clock" ticking slower 
for longer genomes. In this series, the minimum reproduction age was 
proportional to $L$. The most critical was the birthrate which was set for $1$ 
per MC step. Under such a parameter the reproduction potential was regulated 
only by Verhulst factor which controls the population size by killing the 
newborns, and the value of Verhulst factor could be considered a real birthrate 
regulation \cite{Jorge1}. In fact, instead of the birthrate parameter, we have 
the reproduction potential of the population which is an output of the model 
rather than its parameter. 

The results of this third series of
simulations should be discussed separately for the part of the 
genome expressed before the minimum reproduction age (lets call them the
housekeeping genes), and the part expressed after the minimum reproduction age, 
i.e. during the ageing period (death genes) \cite{Kurdziel}. 
One of the measures of the genetic status of genomes or populations is the 
genetic load,  the fraction or frequency of defective genes in the genome. From 
the point of view of the evolutionary costs, the winning strategy is the 
strategy which eliminates more defects by one genetic death. Simulations show 
that under the constant mutational pressure per genome, the proportional 
increase of the number of housekeeping genes with the genome size is associated 
with lowering the fraction of defective housekeeping genes (Figs. 1a and 1b), 
while the number of defective genes eliminated by a single genetic death grows 
with the genome length (Table 2). We would like to stress that the results of 
simulations with haploid genomes or with diploid genomes but with declared 
dominance of defective genes would be different. In such simulations, the number
of defective genes among the housekeeping genes is rather constant and set by 
the threshold $T$ parameter. 

One of the important features of populations with 
larger genomes is their higher reproduction potential. As a result, the 
populations' size grows and the limitations set by the Verhulst factor are 
stronger. One can expect further amelioration of the genetic pool of populations
with larger genomes if the random death introduced by Verhulst factor is 
replaced by selection mechanisms.
The results of simulations, shown as a fraction of populations at a given age 
are shown in Fig. 2. The age axis co-ordinates correspond to the length of the 
genomes and the consecutive genes switched on. Since the numbers of genes in the
genomes and the minimum reproduction age increases, the life span of organisms 
is increasing on this bit scale, too. The plots in Fig. 3 show the age distribution 
of the populations shown in Fig. 2 after normalization of the y-axis scale. 
Note, that the time scale in the plots is still in MC steps. Fig. 4 shows the 
results of normalization when the mutational clock is the base of the time scale,
which means that the x-axis is reduced to the scale corresponding to the 
frequency of mutations. In this figure, the plots representing the age structure
of all simulations give similar results. One can conclude that the Penna model 
has scaling properties when simulations are performed under specifically related
simulation parameters. These results are in better agreement with scaling
than those obtained by 
\cite {Jorge}. Nevertheless, it is important to note that populations with 
different sizes of the genomes and very similar age distributions are 
characterised by very different other parameters describing their genetic 
status as well as reproduction potential. The values of the parameters 
describing the populations are shown in Table 2. 

\section{Conclusions}

It is possible to generate similar age distributions of populations with 
different sizes of genomes of their individuals. Thus the properties of the 
model have become independent of the choice of the Monte Carlo time step.
The main feature of populations
with larger genomes is the higher fidelity of their genome replication: it has 
to be constant per genome rather than per length unit. Populations with larger 
genomes are characterised by higher reproduction potential which results in better 
filling the available environment. The fractions of defective genes among the 
genes indispensable for reaching the reproduction age in larger genomes are 
lower but the number of genes eliminated from the genetic pool by one genetic 
death is higher. The last property is true only for diploid organisms and the 
fraction of loci where the defects are recessive. Haploid, asexually reproducing
organisms have to keep the same number of defects in this part of the genome.


We wish to thank the European project COST-P10 for supporting a 
visit of S. Cebrat at Cologne University. A.{\L}. and S.C. were supported by 
Foundation for Polish Science. D.S. thanks J.S. S\'a Martins for many 
discussions during his unsuccessful attempts to get scaling.

\begin{table}[ht]
\begin{center}
\begin{tabular}{c|c|c|c|c|c} \hline
$L$         &  $R$  &          $M$ &   $B$ &   $C$ &   $T$ \\
32          &   8        &      1  &    1  &    0.015 & 3 \\
64          &   8        &      2  &    1  &    0.03  & 3 \\
128         &   8        &      4  &    1  &    0.06  & 3 \\
256         &   8        &      8  &    1  &    0.12  & 3 \\
512         &   8        &      16 &    1  &    0.24  & 3 \\ \hline \hline
32          &   8        &      1  &    1  &    0.015 & 3 \\
64          &   16       &      2  &    1  &    0.03  & 3 \\
128         &   32       &      4  &    1  &    0.06  & 3 \\
256         &   64       &      8  &    1  &    0.12  & 3 \\
512         &   128      &      16 &    1  &    0.24  & 3 \\ \hline \hline
32          &   8        &      1  &    1  &    0.015 & 3 \\
64          &   16       &      1  &    1  &    0.03  & 3 \\
128         &   32       &      1  &    1  &    0.06  & 3 \\
256         &   64       &      1  &    1  &    0.12  & 3 \\
512         &   128      &      1  &    1  &    0.24  & 3 \\ \hline
\end{tabular}
\end{center}
\caption{
Parameters for three series of simulations, where $M=$ mutation rate per genome.
}
\end{table}

\begin{table}[ht]
\begin{center}
\begin{tabular}{c|c|c|c|c|c|c} \hline
$L$  & $R$   &  maxgen&defects& $V$   & Pop  &  Eliminated \\
32   &  8    &  15    & 0.387 & 0.443 & 5568 &  6.2 \\
64   &  16   &  25    & 0.283 & 0.304 & 6963 &  9.0 \\
128  &  32   &  48    & 0.198 & 0.188 & 8124 &  12.7 \\
256  &  64   &  94    & 0.139 & 0.095 & 9050 &  17.8 \\
512  &  128  &  202   & 0.093 & 0.040 & 9603 &  23.9 \\ \hline
\end{tabular}
\end{center}
\caption{
Characteristics of the populations for third series.
$L$ and $R$ - parameters of simulations; maxgen - the first locus, where
all bits in the genetic pool are set for $1$; defects - average fraction
of defective genes in the section of housekeeping genes; $V$ - fraction  of
surviving newborns (since the birthrate in all simulations was set to $B=1$
per MC step, in fact, Verhulst factor controlled the reproduction rate);
Pop - size of populations in equilibrium, ${N_{max}=10000}$; Eliminated -
average number of defects in the housekeeping genes of one genome.
}
\end{table}

\bigskip
\begin{figure}[hbt]
\begin{center}
\includegraphics[angle=-90,scale=0.5]{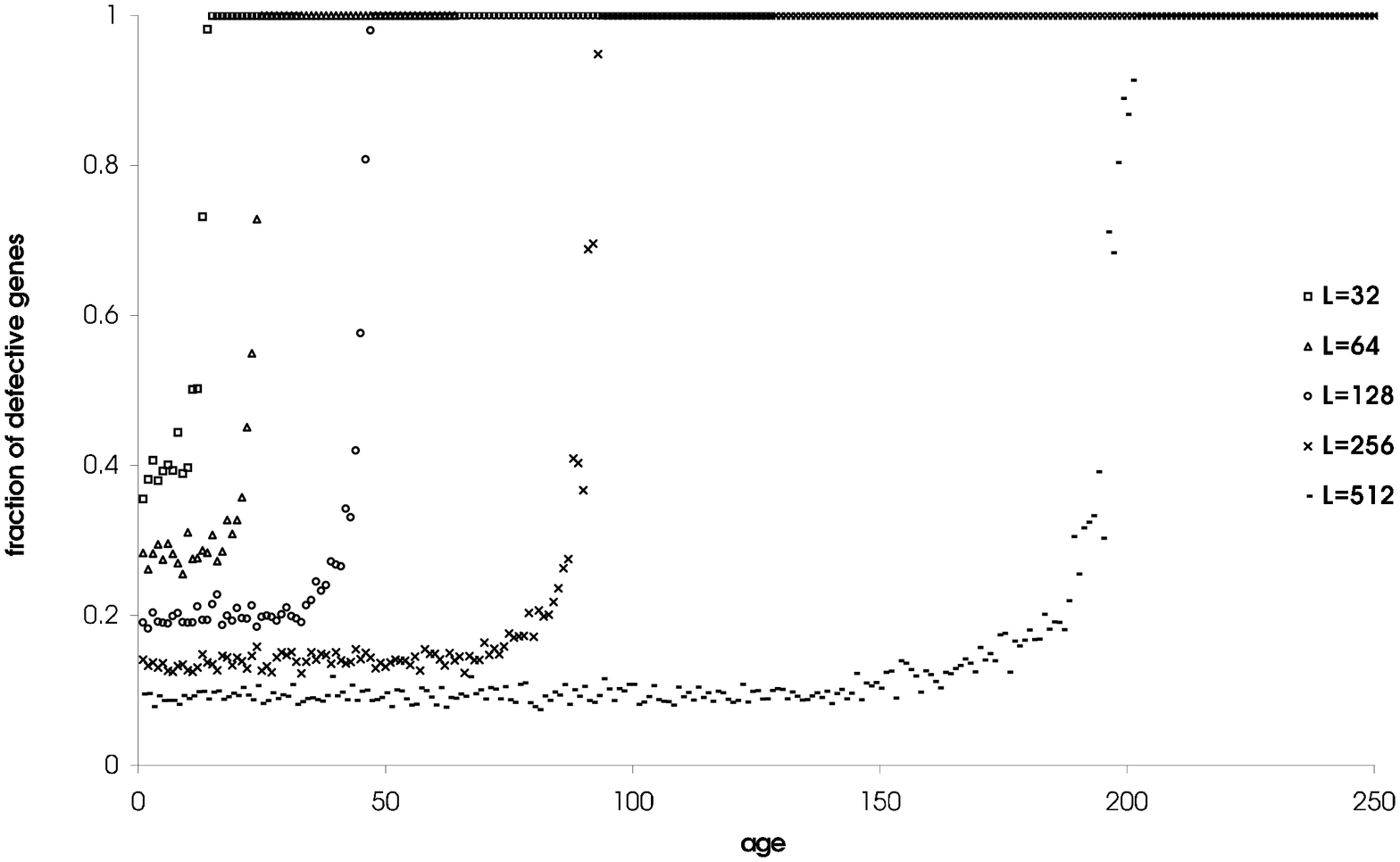}
\includegraphics[angle=-90,scale=0.5]{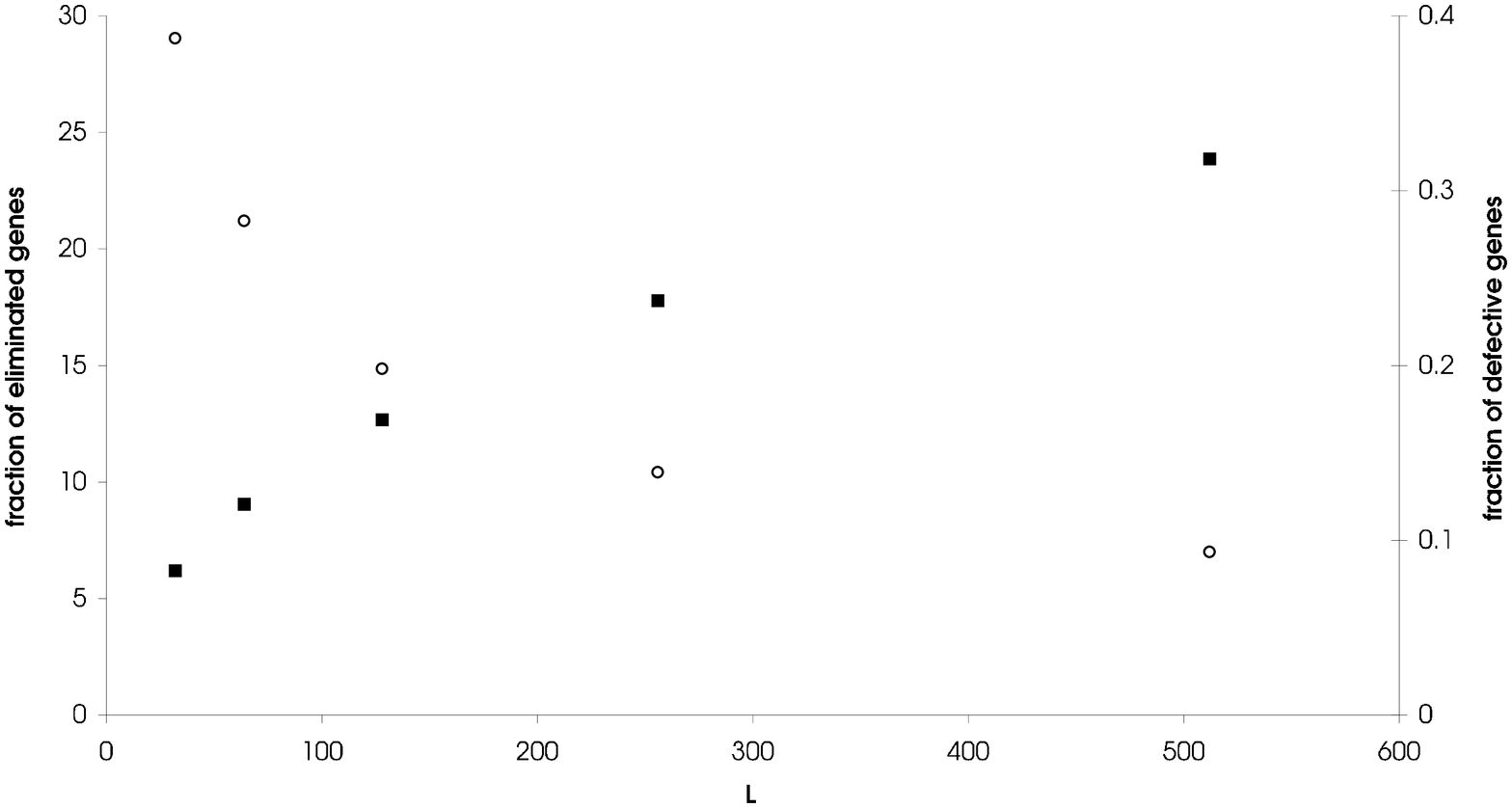}
\end{center}
\caption{
Part a: Distribution of defective genes in the genomes of different length. 
Simulation 
parameters as shown in Table 1 for third series.  X-axis co-ordinates correspond
to the number of bits in the bitstring. 
Part b: Fraction of defective genes in the sections of housekeeping genes in 
genomes of different length (right scale, empty circles); average number of 
defective house keeping genes 
in diploid genomes, genes eliminated by one "genetic death" (left axis, filled 
squares).
}
\end{figure}

\begin{figure}[hbt]
\begin{center}
\includegraphics[angle=-90,scale=0.5]{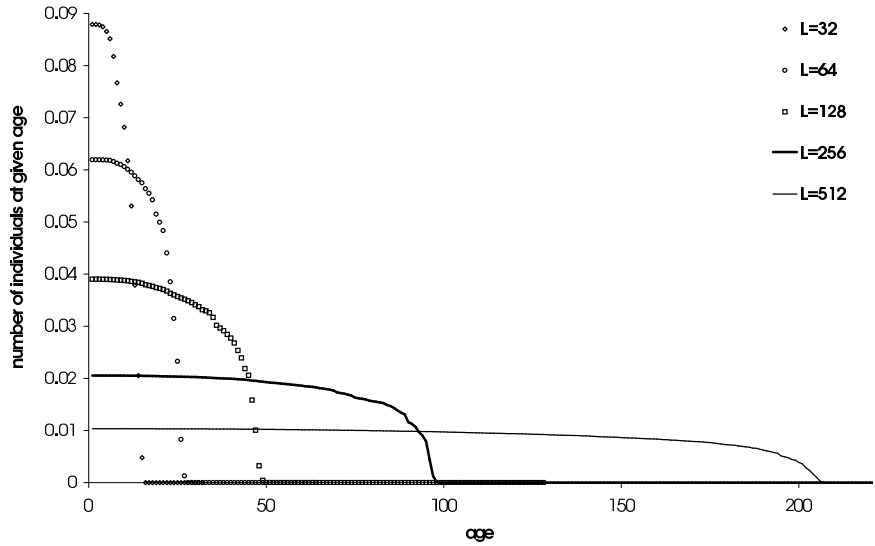}
\end{center}
\caption{
Age distribution of populations with different length of genomes; y-axis
shows fractions of populations at a given age.
}
\end{figure}

\begin{figure}[hbt]
\begin{center}
\includegraphics[angle=-90,scale=0.5]{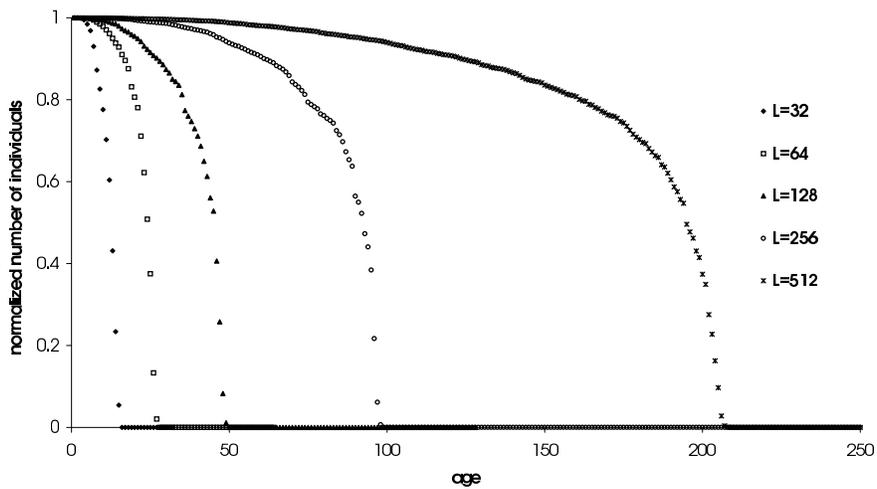}
\end{center}
\caption{
Normalized age distribution of populations. Age scale still in MC steps 
or numbers of bits in  the bitstrings.
}
\end{figure}

\begin{figure}[hbt]
\begin{center}
\includegraphics[angle=-90,scale=0.5]{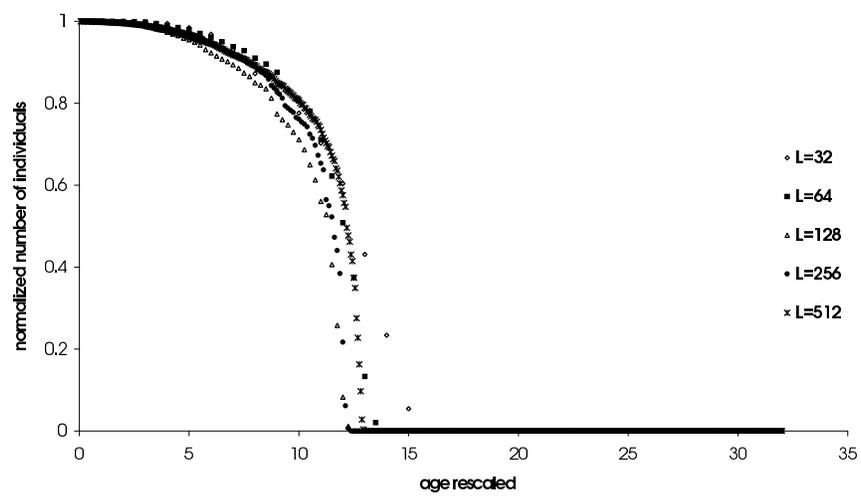}
\end{center}
\caption{
The same as in Fig. 3 but the age axis is rescaled according to the 
mutational pressure  (see text for detailed explanation).  
}
\end{figure}

\end{document}